\documentclass[aip,jcp,reprint]{revtex4-1}
\usepackage{graphicx}
\usepackage{amsmath,amsxtra,amssymb,latexsym, amscd,color}
\usepackage[colorlinks=true, linkcolor=blue, urlcolor=blue, citecolor=blue]{hyperref}

\begin{document}

\title{Protein escape at the ribosomal exit tunnel: Effect of
the tunnel shape}

\author{Phuong Thuy Bui}
\affiliation{Institute of Theoretical and Applied Research, Duy Tan University,
Hanoi, 100000, Vietnam}
\affiliation{Faculty of Pharmacy, Duy Tan University, Da Nang, 550000,
Vietnam}

\author{Trinh Xuan Hoang}
\email[Corresponding author, E-mail: ]{hoang@iop.vast.vn}
\affiliation{Institute of Physics, Vietnam Academy of Science and Technology,
10 Dao Tan, Ba Dinh, Hanoi 11108, Vietnam}
\affiliation{Graduate University of Science and Technology, Vietnam Academy of
Science and Technology, 18 Hoang Quoc Viet, Cau Giay, Hanoi 11307, Vietnam}

\date{\today}

\begin{abstract}
We study the post-translational escape of nascent proteins at the ribosomal
exit tunnel with the consideration of a real shape atomistic tunnel based on
the Protein Data Bank (PDB) structure of the large ribosome subunit of archeon
{\it Haloarcula marismortui}. Molecular dynamics simulations employing the
Go-like model for the proteins show that at intermediate and high
temperatures, including a presumable physiological temperature, the protein
escape process at the atomistic tunnel is quantitatively similar to that at a
cylinder tunnel of length $L=72$~{\AA} and diameter $d=16$~{\AA}. At low
temperatures, the atomistic tunnel, however, yields an increased
probability of protein trapping inside the tunnel while the cylinder
tunnel does not cause the trapping. All-$\beta$ proteins tend to escape faster
than all-$\alpha$ proteins but this difference is blurred on increasing the
protein's chain length. A 29-residue zinc-finger domain is shown to be severely
trapped inside
the tunnel. Most of the single-domain proteins considered, however, can escape
efficiently at the physiological temperature with the escape time
distribution following the diffusion model proposed in our previous works.  
An extrapolation of the simulation data to a realistic value of the friction
coefficient for amino acids indicates that the escape times of
globular proteins are at the sub-millisecond scale.
It is argued that this time scale is short enough for the smooth functioning
of the ribosome by not allowing nascent proteins to jam the ribosome tunnel.
\end{abstract}

\maketitle

\section{Introduction}

After the determination of the ribosome structures
\cite{Gabashvili2000,Ban2000}, there has been growing attention on
understanding the role of the ribosomal exit tunnel on protein biosynthesis and
on cotranslational protein folding (for recent reviews see Ref.
\cite{Cavagnero2011,Obrien2016,Javed2017,Rodnina2017}). 
Biochemical studies indicate that the tunnel plays an active role in the
regulation of the protein translation process by blocking specific peptide
sequences \cite{Ito2002,Tenson2002}. The mechanism of this blocking or
translation arrest can be associated with certain ribosomal protein's motion
which alters the tunnel shape \cite{Berisio2003}. 
In cotranslational protein folding, the tunnel imposes a spatial
confinement on the traversing nascent peptide
while it is being synthesized by the ribosome.
The narrow geometry of the tunnel was suggested to entropically promote the
$\alpha$-helix formation \cite{Thirumalai2005} whereas it may sterically
obstruct the formation of $\beta$-sheet \cite{Makarov2004}. 
Depending on the location within the tunnel, 
peptides can form simple $\alpha$-helices \cite{Deutsch2005} and small
tertiary structure units \cite{Deutsch2009}. The latter can be observed near
the tunnel exit port where there is enough space to hold the structure.
Simulations \cite{Elcock2006,OBrien2010} and experiments
\cite{Holtkamp2015,Nilsson2015,Kudva2018} indicate that cotranslational folding
starts inside the tunnel, with the structures ranging from a non-native
compact conformation \cite{Holtkamp2015} and transient tertiary structures
\cite{OBrien2010} to a small protein domain \cite{Nilsson2015}.
There are also considerations that the folding inside the ribosome tunnel is
negligible leading to a focus only on the folding of nascent chains as they
emerge from the tunnel, as shown in studies with stalled ribosome-bound nascent
chain experiments \cite{Cabrita2009,Eichmann2010} and simulations
\cite{OBrien2011,Shakh2013}.

In recent works \cite{Thuy2016,Thuy2018}, we suggested that the exit tunnel, as
a passive conduit, has a significant impact on the early
post-translational folding, i.e. shortly after the protein's C-terminus is
released from the peptidyl transferase center (PTC). This
impact corresponds to a vectorial folding \cite{Thuy2016} induced by the tunnel
and is associated with the escape process of a full length protein from the
tunnel. In particular, the folding and escape of a nascent protein at the
tunnel are concomitant with each other. Folding accelerates the escape process
whereas a gradual escape improves the protein foldability.
Furthermore, we showed that the protein escape at the tunnel is governed by 
a diffusion mechanism and the escape time distribution can be captured
by a simple model of a Brownian particle in a linear potential field.

The escape process also has an important meaning of itself. It should not be
too quickly because this would leave an escaped protein significantly
unfolded outside the ribosome, by that increases the chance of protein aggregation
\cite{Dobson2003}. It cannot be also too slow because this
would decrease the productivity of the ribosome. Interestingly, our previous
study shows that the real length of ribosome exit tunnel is close to a
cross-over tunnel length \cite{Thuy2018} of 90 to 110~{\AA} for the diffusion
of small globular proteins. For tunnels of lengths larger than this cross-over
length, the diffusion is much slower. Thus, it was suggested that the ribosome
tunnel length may have been selected by evolution to facilitate an appropriate
escape time.

The previous works \cite{Thuy2016,Thuy2018} on the protein escape process
considered a highly simplified model of the ribosomal exit tunnel, being
a hollow cylinder with repulsive wall. Real exit tunnel is highly porous for
water-size molecules and effectively adopts an irregular shape \cite{Voss2006}
for polypetides. The tunnel shape also depends on the type of organism
\cite{Khanh2019}. The aim of our present study is to work with
a realistic tunnel to test the validity of the previous findings, and to
investigate the effect of the tunnel shape on the protein escape process. For
this purpose, we consider an atomistic model of the tunnel based on the
resolved PDB structure of the large ribosome subunit of {\it Haloarcula
marismortui} \cite{Ban2000}. The atomistic tunnel incorporates all heavy
atoms for the ribosomal RNA but only the $C_\alpha$'s for the ribosomal
proteins. The $C_\alpha$-only representation is also used for nascent
proteins within a standard Go-like model
\cite{Go1983,HoangJCP2000,HoangJCP2000b,Clementi2000}. We have chosen the
same coarse-grained level for ribosomal proteins for a
consistency in modelling with the chosen Go-like model. This choice,
however, reduces the roughness of the tunnel's wall where amino acid
side-chains are exposed. In order to
accelerate the simulations all the ribosomal atoms are kept fixed, thus, the
tunnel acts solely as a passive channel for the escaping proteins in our
consideration.

The main focus of our paper is on the effect of the tunnel shape on the escape
process. In order to delineate this effect, we compared the escape of a nascent
protein at the atomistic tunnel with that at an equivalent cylinder tunnel. The
latter is described such that it produces a similar median escape time to
that obtained with the atomistic tunnel. This comparison shows similarities and
differences between the two tunnels. Remarkably, we find that the atomistic
tunnel yields a non-zero probability of trapping protein inside the tunnel
while the cylinder tunnel does not.  Other issues being discussed are the
dependences of the escape time on temperature, on protein's length and native
state topology among small single domain proteins, and on the friction
coefficient for amino acids.  An estimation of real escape time from the
simulations is also presented.
Interestingly, we find that the estimated escape time is relevant to the
functionality of the ribosome.

\section{Models and Methods}

\subsection{Protein and tunnel models}

Nascent proteins are considered in a Go-like model
\cite{Go1983,HoangJCP2000,HoangJCP2000b,Clementi2000}, in which a protein
is represented only by its C$_\alpha$ atoms.  The intramolecular potential
energy of a protein in a conformation is given by \cite{Clementi2000}
\begin{eqnarray}
E & = & \sum_{i=1}^{N-1} K_b (r_{i,i+1} - b)^2 +
\sum_{i=2}^{N-1} K_\theta (\theta_i - \theta_{i}^*)^2 + \nonumber \\
& &
+ \sum_{n=1,3} \sum_{i=2}^{N-2} K_\phi^{(n)} [1+\cos(n(\phi_i - \phi_{i}^*))] + 
\nonumber \\
& &
+ \sum_{i+3<j}
\epsilon \left[ 5\left( \frac{r_{ij}^*}{r_{ij}} \right)^{12} 
- 6 \left( \frac{r_{ij}^*}{r_{ij}} \right)^{10} \right] \Delta_{ij} 
+\nonumber \\
& &
+ \sum_{i+3<j}
\epsilon \left( \frac{\sigma}{r_{ij}} \right)^{12} (1-\Delta_{ij}) \ ,
\label{eq1}
\end{eqnarray}
where $N$ is the number of amino acid residues; $r_{ij}$ is the distance
between residues $i$ and $j$; $\theta_i$ and $\phi_i$ are the bond and
dihedral angles of the residue $i$; $n$ is equal to either 1 or 3; the star
superscript corresponds to the native state's value; $\Delta$ is the native
contact map with
$\Delta_{ij}$ equal to 1 if there is a native contact between $i$ and $j$ and
equal to 0 otherwise. $\Delta$ is defined based on an all-atom consideration of
the protein PDB structure with a contact cut-off distance between two atoms
equal to 1.5 times the sum of their
atomic van der Waals (vdW) radii (the C3 map in Ref. \cite{Thuy2018}).
Energy is given in units of $\epsilon$, which corresponds to the depth of the
10-12 Lennard-Jones potential given in the fourth term of Eq. (\ref{eq1}).
The parameters used in the model are $b=3.8$~{\AA},
$\sigma=5$~{\AA}, $K_b=100~\epsilon${\AA}$^{-2}$,
$K_\theta=20~\epsilon(\mathrm{rad})^{-2}$, $K_\phi^{(1)}=-\epsilon$,
$K_\phi^{(3)}=-0.5~\epsilon$.

\begin{figure}
\center
\includegraphics[width=8.5cm]{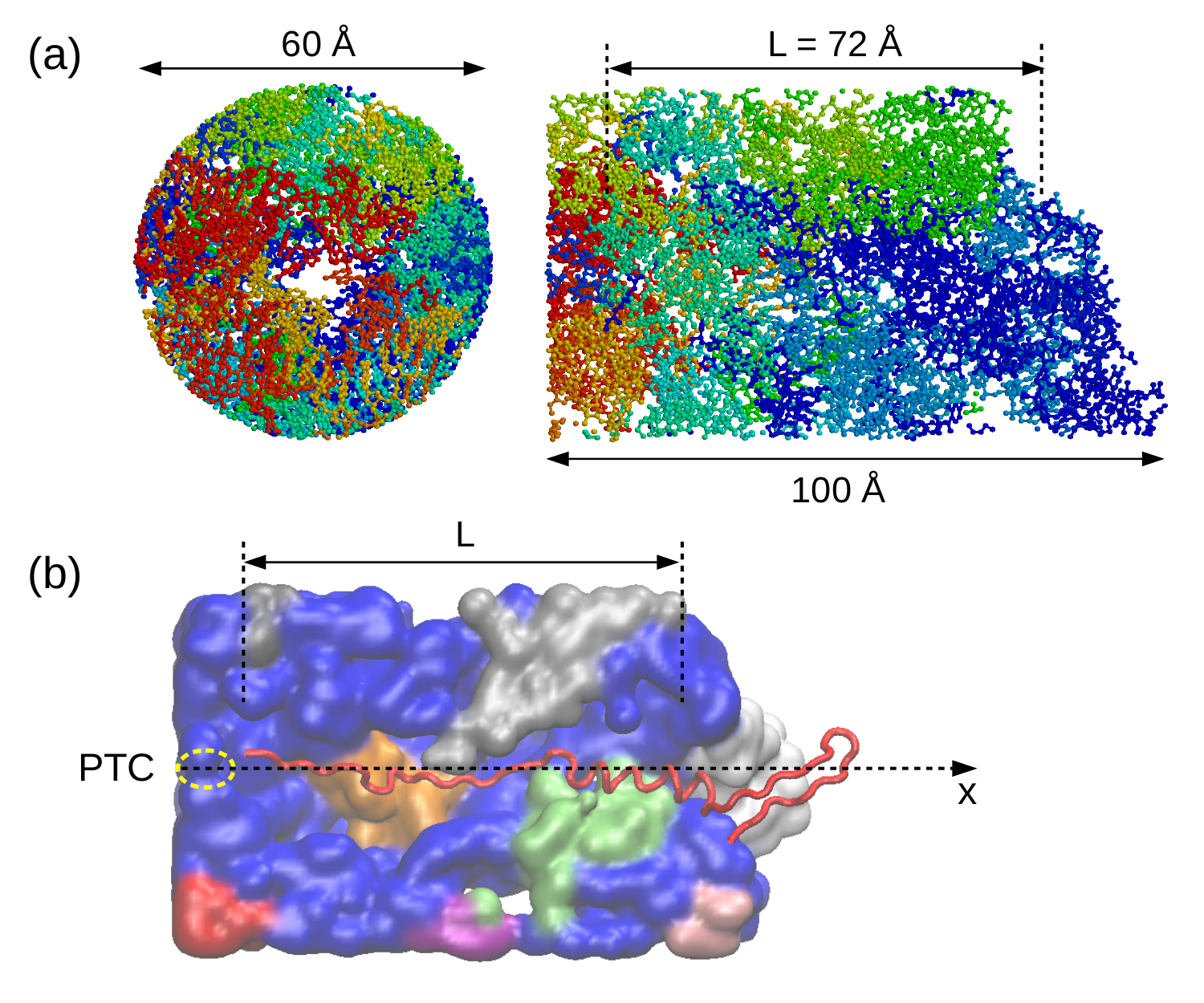}
\caption{(a) Two projected views of the ribosomal exit tunnel of {\it H.
Marismortui} represented by all heavy atoms within 30 {\AA} away from a
chosen axis that goes roughly through the middle of the tunnel. 
(b) A conformation of protein GB1 (red) obtained
by simulation after the protein is grown from the PTC at the
atomistic tunnel.  The latter is shown from a cross-section plane which
includes the tunnel axis. Assuming that $x$ is the tunnel axis, the planes of
the projected views are $y$-$z$ and $x$-$y$ in (a) and $x$-$z$ in (b).
The PTC region is identified as located between the
bases A2486, U2620 and C2104 of the ribosomal RNA \cite{Steitz2005}. 
The tunnel length of $L=72$~{\AA} indicated by arrows corresponds to an
axial distance from the opening of the tunnel near the PTC to an inner
edge of the tunnel exit port. 
} 
\label{fig:tunnel}
\end{figure}

To build up the atomistic model of the ribosomal exit tunnel, we used the 
crystal structure of the {\it Haloarcula Marismortui}'s large ribosomal
subunit with the PDB code 1jj2 \cite{Ban2000}. Only a part of the 
subunit surrounding the tunnel was taken to the model (Fig. \ref{fig:tunnel}a).
In particular, we excluded the ribosome's atoms that are further than 30 {\AA}
from a chosen axis $x$ that originates from the PTC region and goes
roughly through the middle of the tunnel. Furthermore, we kept only heavy atoms
for the ribosomal RNA and only C$_\alpha$ atoms for the ribosomal proteins.
The interaction potential between the C$_\alpha$ atoms of a
nascent
protein and the C$_\alpha$ atoms of a ribosomal protein is assumed to be
repulsive and takes the form of $\epsilon(\sigma/r_{ij})^{12}$. 
The interaction potential between a nascent protein's C$_\alpha$ ($a$) and a
ribosomal RNA heavy atom ($b$) is also repulsive and given by
\begin{equation}
V(r) = \epsilon \left( \frac{R_a + R_b + R_+}{r_{ab}} \right)^{12} \ ,
\end{equation}
where $r_{ab}$ is the center-to-center distance between $a$ and $b$; $R_a =
2.5$ {\AA} is an effective radius of amino acid; $R_b$ is the atomic vdW radius
of
the ribosomal atom; $R_{+}=0.8$~{\AA} is an effective additive radius
accounting for the fact that hydrogen atoms are not considered in the model of
the tunnel. For example, if $b$ is a carbon atom (with a vdW radius of
1.7~{\AA}) then one gets $R_b + R_+ = 2.5$~{\AA}, i.e. same as $R_a$.
We have checked that the above mentioned value of $R_+$ yields an adequate
escape behavior for the nascent proteins \cite{Thuyjpcs2020}. 

The simulations were carried out using molecular dynamics method based on the
Langevin equation of motion and a Verlet algorithm \cite{Thuy2016}. The amino
acids are assumed to have an uniform mass, $m$. Temperature is given in units
of $\epsilon/k_B$, whereas time is measured in units of
$\tau=\sqrt{m\sigma^2/\epsilon}$. We used the same value of the friction
coefficient in the Langevin equation, $\zeta = 1~m \tau^{-1}$, through out
the studies except in the Subsection III.C where the dependence on
the friction coefficient is investigated. In each simulation, first the
polypeptide chain was grown in the tunnel from the PTC at a
constant speed given by the growth time $t_g = 100~\tau$ per amino acid. 
This growth time is slow enough to produce converged properties of fully
translated protein conformations in terms of the radius of gyration and the
number of native contacts, i.e. the distributions of these quantities are
similar to those obtained with a much larger growth time \cite{Thuy2016}.
After the protein is completely translated, the simulation was run until it has
fully escaped from the tunnel. The escape time was measured from the moment of
complete translation.
Because the exit port is irregular with a complex geometry, in order to capture
the essential escape time, we have defined the tunnel region as the space
within a cylinder of length $L=72$~{\AA} and radius of 15~{\AA} centered about
the tunnel axis $x$.  This region starts from an opening of the tunnel near the
PTC and ends at an
inner edge of the exit port (Fig.~\ref{fig:tunnel}), corresponding to
positions from $x=10$~{\AA} to $x=82$~{\AA}.
An amino acid residue is considered to have escaped from the tunnel if it is found
outside the tunnel region.
Typically, for each
temperature about 1000 independent growth and escape trajectories are simulated
to obtain the statistics of the escape time.

\subsection{Diffusion model}

The escape of a fully translated protein at the ribosome tunnel is
driven by: (i) an enthalpic force associated with the folding of the protein
outside the tunnel, (ii) an entropy gain as the chain
emerges from the tunnel, and (iii) the stochastic motion of a partially
unfolded chain. It has been shown that in the Go-like model, the free energy
change of a protein along an escape coordinate is a monotonically decreasing
function, which is approximately linear at intermediate and high temperatures
\cite{Thuy2016,Thuyjpcs2020}.  
Interestingly, all these effects can be effectively acquired in a
diffusion model \cite{Thuy2016,Thuy2018}, which describes the protein
escape as the diffusion of a Brownian particle pulled by a constant 
force in one dimension.
The particle diffusion in a potential field $U(x)$ is governed by the Smoluchowski
equation \cite{vankampen}
\begin{equation}
\frac{\partial}{\partial t}\, p(x,t) =  
\frac{\partial}{\partial x} D
 \left(\beta \frac{\partial U(x)}{\partial x} + \frac{\partial}{\partial x}
\right)\, p(x,t) ,
\label{eq:smolu}
\end{equation}
where $p(x,t)$ is a probability density of finding the
particle at position $x$ and at time $t>0$, given that it was found 
at position $x=0$ at time $t=0$; $D$ is the diffusion constant, assumed to be
position independent; and $\beta=(k_B T)^{-1}$ is the inverse temperature
with $k_B$ the Boltzmann constant.
The escape time is described as the first passage time of the particle reaching
a distance $L$ from an origin in the drift direction. Given an external
potential of the linear form $U(x)=-kx$, where $x$ is the coordinate of the
particle and $k$ is the force, the
distribution of the escape time was obtained via an exact solution
\cite{Cox} and is given by \cite{Thuy2018}
\begin{equation}
g (t)=\frac{L}{\sqrt{4\pi D t^3}} \exp\left[
-\frac{(L - D\beta k t)^2}{4Dt}
\right] .
\label{eq:gt}
\end{equation}
Using the distribution in Eq. (\ref{eq:gt}) one obtains the mean escape
time
\begin{equation}
\mu_t \equiv \langle t \rangle = \int_0^\infty t \, g(t)\, dt 
= \frac{L}{D \beta k} \ ,
\label{eq:mut}
\end{equation}
with $D \beta k$ as the mean diffusion speed, 
and the standard deviation
\begin{equation}
\sigma_t \equiv \sqrt{\langle t^2 \rangle - \langle t \rangle^2}
= \frac{\sqrt{2 L}}{D (\beta k)^{\frac{3}{2}}} \ .
\label{eq:sigt} \end{equation}
Note that both $\mu_t$ and $\sigma_t$ diverges when $k=0$, for which $g(t)$
becomes a heavy-tailed L\'evy distribution.
It has been shown that $D$ and $\beta k$ depend on $L$, on the protein and on
other conditions such as the crowders' volume fraction outside the tunnel
\cite{Thuy2018}.

\section{Results and Discussion}

\subsection{Effect of tunnel shape on escape process}

To study the effect of tunnel shape on the escape process, we consider the
immunoglobulin binding (B1) domain of protein G (GB1) as a nascent protein. 
This protein has a length of $N=56$ amino acids and was considered in our
previous studies \cite{Thuy2016,Thuy2018}. The folding temperature of GB1 was
found as $T_f=1.004~\epsilon/k_B$ \cite{Thuy2018}. Experimentally, the melting
temperature of the wild-type GB1 at pH 5.5 has been reported to be
80.5$^\circ$C \cite{Campos2002}. We will study the escape process at
various temperatures but will focus on the simulation temperature
$T=0.85~\epsilon/k_B$,
which after unit conversion corresponds to a physiologically relevant
temperature of 26.3$^\circ$C.

\begin{figure}
\includegraphics[width=8.5cm]{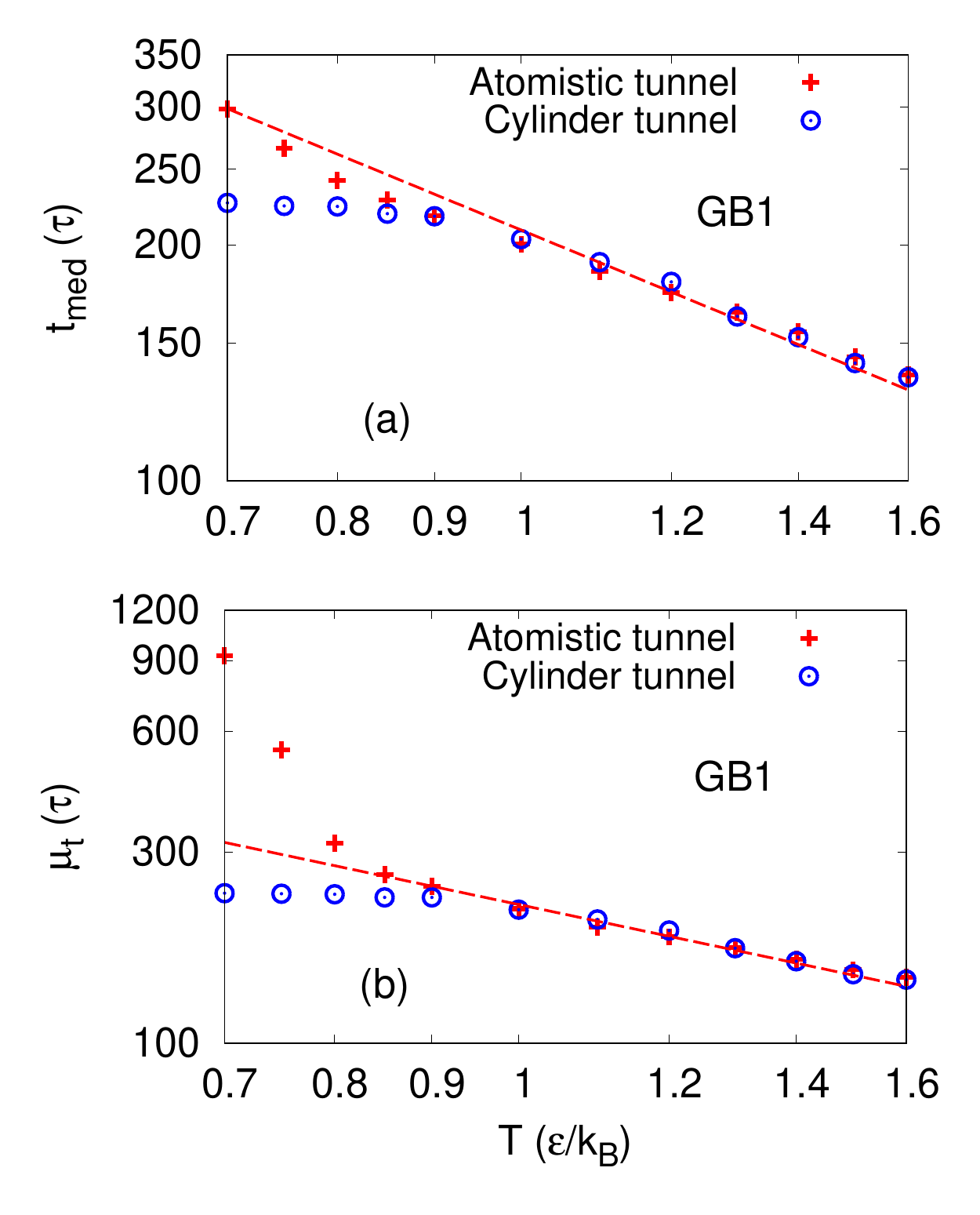}
\caption{Log-log plots of the temperature dependence of the median escape time
$t_\mathrm{med}$ (a) and the mean escape time $\mu_t$ (b) for protein GB1
at the atomistic tunnel (crosses) and at a cylinder tunnel of length
$L=72$~{\AA} and diameter $d=16$~{\AA} (circles). The
straight line (dashed) has a slope of $-1$ and is fitted to the data
points of the cylinder tunnel at high temperatures ($T\ge 1~\epsilon/k_B$).
}
\label{fig:tmedlog}
\end{figure}

In order to delineate the effects of tunnel shape on the escape process
we sought for an equivalent cylinder tunnel that yields a similar 
escape time to that obtained at the
atomistic tunnel for the protein GB1. We found that among the cylinder
tunnels of the same length as that of the atomistic tunnel, i.e.  $L=72$~{\AA},
the one with diameter $d=16$~{\AA} satisfies quite well the last requirement
over a wide range of temperature.
Figure \ref{fig:tmedlog}(a) shows that the median escape times $t_\mathrm{med}$
for the atomistic tunnel and the cylinder tunnel are quite close to each other
for various temperatures from $0.85~\epsilon/k_B$ to $1.6~\epsilon/k_B$.
Figure \ref{fig:tmedlog}(b) also shows that the mean escape times $\mu_t$ for
the two tunnels agree very well with each other at intermediate and
high temperatures ($T > 0.85~\epsilon/k_B$). For $T \leq 0.85~\epsilon/k_B$,
both $t_\mathrm{med}$ and $\mu_t$ for the atomistic tunnel are larger than for
the cylinder tunnel and the differences increase as the temperature decreases.
These differences indicate that at low temperatures, it is more difficult
for proteins to escape from the atomistic tunnel than from the cylinder one. It
appears that the physiological temperature $0.85~\epsilon/k_B$ corresponds to a
borderline behavior of the escape process, in which the effect of tunnel shape
starts to get in.

Figure \ref{fig:tmedlog} also shows the linear dependences of $t_\mathrm{med}$
and $\mu_t$ on $T^{-1}$ in log-log scales (the dashed lines) as it would be found
for a Brownian particle diffused in a potential field with a constant $\beta k$. 
Our previous study \cite{Thuy2018} shows that this linear behavior is found for
a homopolymer chain with self-repulsion, and thus can be applied for
intrinsically disordered proteins. For foldable proteins like the GB1, this
linear dependence can be observed only at high temperatures, at which the 
proteins are unfolded during the escape.

Note that one can also have an equivalent cylinder tunnel of a length
different from that of the atomistic tunnel. For example, we found that 
the cylinder tunnel of $L=82$~{\AA} and $d=13.5$~{\AA} also yields similar
median and mean escape times to those of the atomistic tunnel (see Fig.~S1
of the supplementary material)
with an equally good agreement as in Fig. \ref{fig:tmedlog}. For the most 
relevant comparison, we consider only the equivalent cylinder tunnel
of $L=72$~{\AA} and $d=16$~{\AA}.

\begin{figure*}
\includegraphics[width=17cm]{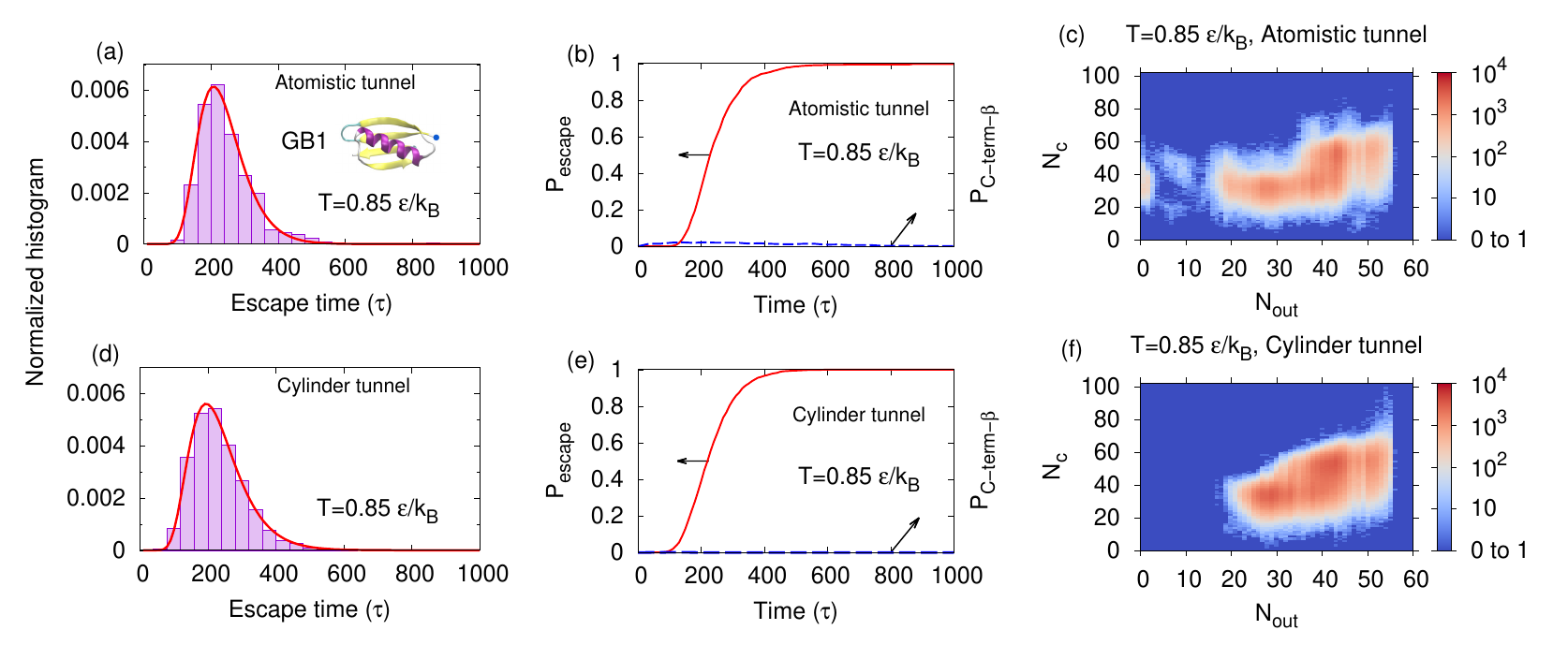}
\caption{Distributions of the escape time (a,d), time dependences of the escape
probability $P_\mathrm{escape}$ (solid) and the probability of C-terminal
$\beta$-hairpin formation inside the tunnel $P_{\textrm{C-term-}\beta}$
(dashed) (b,e), and histograms of conformations as a function of the number of
residues escaped from the tunnel $N_\mathrm{out}$
and the number of native contacts $N_c$ (c,f) for protein GB1 at the
atomistic tunnel (upper panels) and at an equivalent cylinder tunnel (lower
panels) of length $L=72$~{\AA} and diameter $d=16$~{\AA}, at temperature
$T=0.85~\epsilon/k_B$. The native conformation of GB1, shown in (a) as
inset, has 120 native contacts.
}
\label{fig:compare}
\end{figure*}

We now examine more carefully the escape processes of GB1 at the atomistic
and cylinder tunnels at $T=0.85~\epsilon/k_B$.
Figure \ref{fig:compare}(a) shows that at the atomistic tunnel, the histogram
of the escape time for this protein obtained from the simulations follows quite
well the distribution function $g(t)$ given by the diffusion model in Eq.
(\ref{eq:gt}).
Figure \ref{fig:compare}(b) shows that the probability of protein escape
$P_\mathrm{escape}$ has a sigmoidal dependence on the time $t$ with
$P_\mathrm{escape}$ reached the value of 1 at $t\approx 900~\tau$. This result
means that the protein can effectively escape from the tunnel without 
significant delays compared to the median escape time, $t_\mathrm{med}
\approx 230~\tau$. Fig. \ref{fig:compare}(b) also shows the dependence
of the probability $P_{\textrm{C-term-}\beta}$ of forming the C-terminal
$\beta$-hairpin inside the tunnel on time. The time dependence
of this probability is obtained by averaging over multiple escape trajectories. 
The C-terminal $\beta$-hairpin is said to be formed inside the tunnel if it
forms at least half of its native contacts and when all of its residues (41-56)
are located within the tunnel.
We tracked this
$\beta$-hairpin because previous study showed that at low temperatures the GB1
protein can escape from a cylinder tunnel through two different pathways
depending on whether the C-terminal $\beta$-hairpin is formed inside the tunnel
or not \cite{Thuy2016}. 
Fig. \ref{fig:compare}(b) shows that at the atomistic tunnel, only 
a small fraction of about 2\% of the escape trajectories have this
$\beta$-hairpin formed inside the tunnel. We have checked that the trajectories
having this hairpin formed typically correspond to longer
escape times than other trajectories.

Figure \ref{fig:compare}(c) shows the histogram of conformations observed
during the escape process as a function of the number of native contacts $N_c$
and the number of residues escaped from the tunnel $N_\mathrm{out}$ for
protein GB1 at the atomistic tunnel.
The histogram shows a high-density cloud of conformations having intermediate
values of $N_c$ and $N_\mathrm{out}$, indicating that the protein folds during
the escape. The blurring of the cloud, however, suggests that the protein
adopts a wide range of conformations during the escape process.
Given that the maximum $N_c$ for GB1 is 120, the histogram shows that
during the escape the protein can form up to two-thirds of all of its native
contacts.
Note that conformations of $N_\mathrm{out}=0$, i.e. completely located within
the tunnel, are also present in the histogram.

\begin{figure*}
\includegraphics[width=17cm]{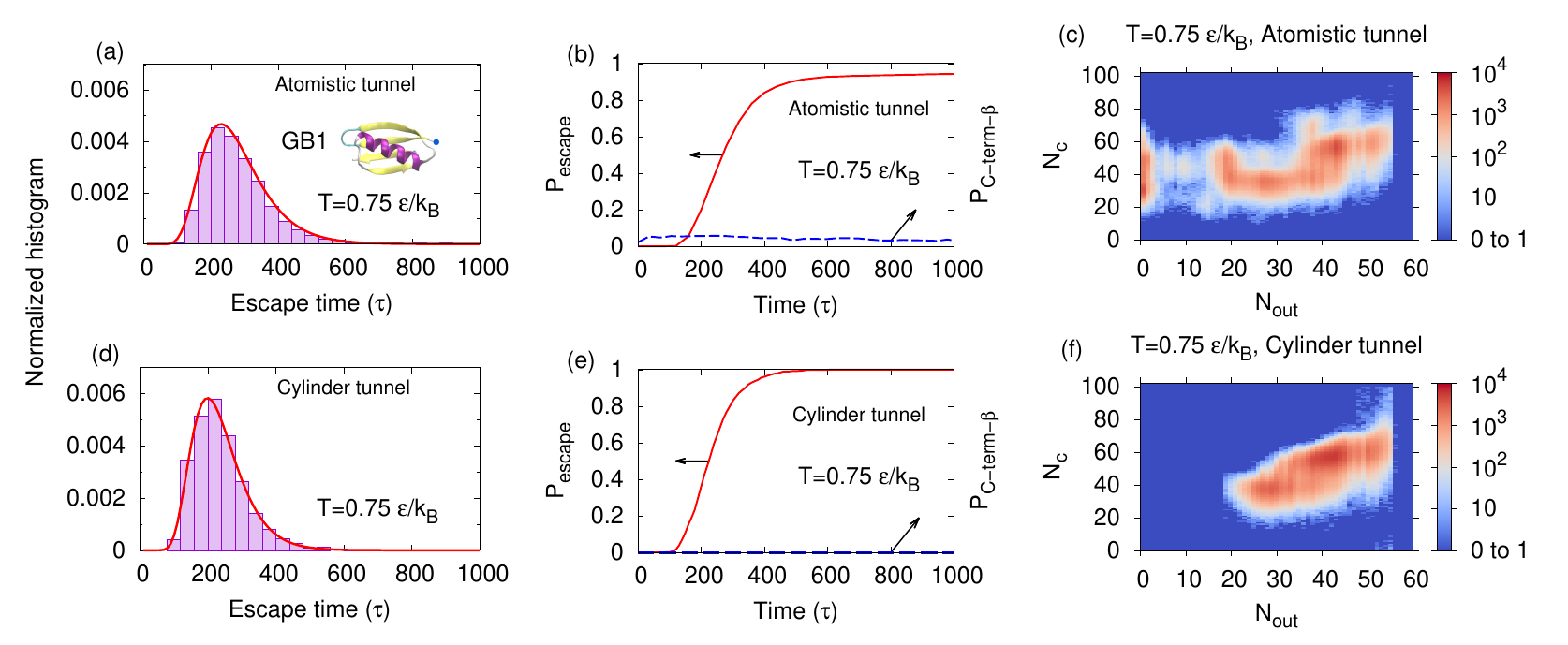}
\caption{Same as Fig. \ref{fig:compare} but for $T=0.75~\epsilon/k_B$.
The panels show
distributions of the escape time (a,d), the time dependences of the escape
probability $P_\mathrm{escape}$ (solid) and the probability of C-terminal
$\beta$-hairpin formation inside the tunnel $P_{\textrm{C-term-}\beta}$
(dashed) (b,e), and the histograms of conformations as a function of the number
of residues escaped from the tunnel $N_\mathrm{out}$
and the number of native contacts $N_c$ (c,f) for protein GB1 at the
atomistic tunnel (upper panels) and at an equivalent cylinder tunnel (lower
panels) of length $L=72$~{\AA} and diameter $d=16$~{\AA}.
}
\label{fig:lowt}
\end{figure*}

\begin{figure*}
\includegraphics[width=15cm]{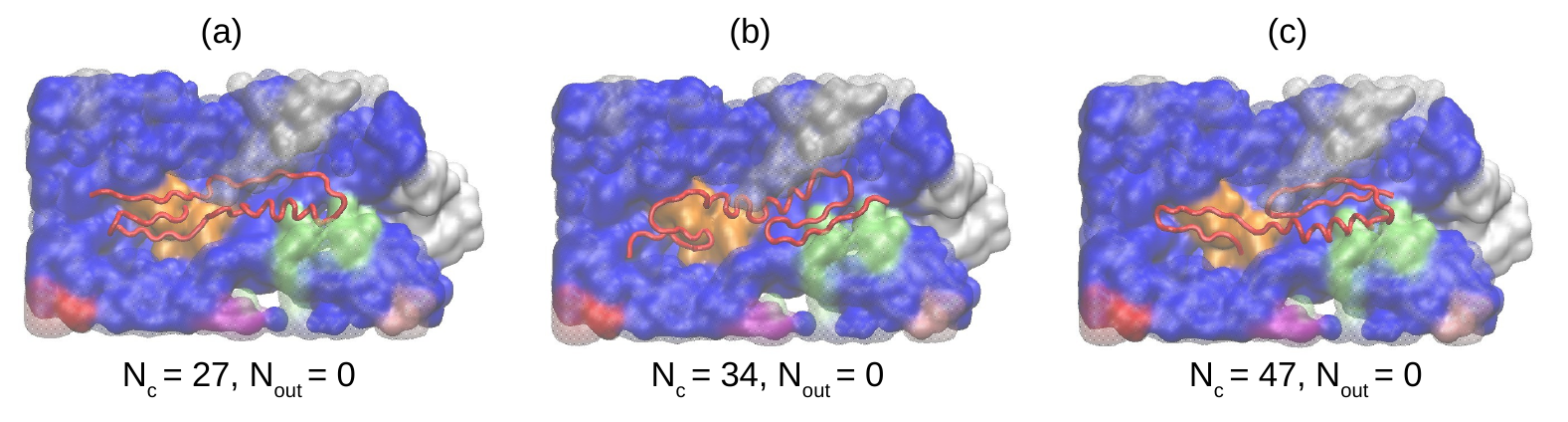}
\caption{Examples of the conformations of GB1 that are trapped inside the
atomistic tunnel during the escape process at $T=0.75~\epsilon/k_B$. The
number of native contacts of the conformations are $N_c=27$ (a), $N_c=34$ (b)
and $N_c=47$ (c), whereas all of them have $N_\mathrm{out}=0$, as indicated.}
\label{fig:trap}
\end{figure*}

Figures \ref{fig:compare}(d,e,f) show that the equivalent cylinder tunnel of
diameter $d=16$~{\AA} produces not only a similar escape time distribution but
also a similar 
dependence of $P_\mathrm{escape}$ on time and a similar histogram of escaping
protein conformations to those obtained with the atomistic tunnel. Notice,
however, that there are differences. First, the escape time distribution
at the atomistic tunnel is slightly more narrow than the one at the cylinder
tunnel while the median escape time at the cylinder is slightly smaller
than at the atomistic tunnel (220~$\tau$ vs. 230~$\tau$).
Second, for the cylinder tunnel the escape probability
$P_\mathrm{escape}$ reaches 1 faster at the time about $500~\tau$.
Third, the
histogram in $N_c$ and $N_\mathrm{out}$ less dispersed in the case of the
cylinder tunnel. These differences indicate that the protein escapes relatively
more easily at the cylinder tunnel than at the atomistic tunnel. 

Other differences at the two tunnels can be seen at the probability of
forming the C-terminal $\beta$-hairpin and the histogram of
conformation during the escape process. Fig. \ref{fig:compare}(e) shows that
for the cylinder tunnel the probability $P_{\textrm{C-term-}\beta}$ is
zero at all times, indicating that the C-terminal $\beta$-hairpin 
does not form inside the cylinder tunnel. 
Fig. \ref{fig:compare}(f) shows that the histogram of conformations during
the escape process for the cylinder tunnel does not include conformations
of small $N_\mathrm{out}$ (less than about 16). 
These results are different to those at the atomistic tunnel and indicate that
the pathways at the atomistic tunnel are more diverse than at the
cylinder tunnel.

The differences between the escape processes at the two tunnels magnify as the
temperature is lowered. In Fig. \ref{fig:lowt} we show the same plots as in
Fig. \ref{fig:compare} but for $T=0.75~\epsilon/k_B$. 
Figures \ref{fig:lowt}(a) and \ref{fig:lowt}(d) show that the escape time
distribution for the atomistic tunnel is significantly more broad than for the
cylinder tunnel. Figures \ref{fig:lowt}(b) and \ref{fig:lowt}(e) show that for
the atomistic tunnel the escape probability $P_\mathrm{escape}$ can reach only
about 94\% at $t=1000~\tau$ while for cylinder tunnel it can reach 100\% at $t
\approx 500~\tau$. 
Figure \ref{fig:lowt}(b) shows that at $T=0.75~\epsilon/k_B$, about 5\% of
the escape trajectories having the C-terminal $\beta$-hairpin formed inside the
atomistic tunnel while this fraction remains to be
zero for the cylinder tunnel (Fig. \ref{fig:lowt}(e)). Figures
\ref{fig:lowt}(c) and \ref{fig:lowt}(f) show that the histogram of escaping
conformations for the atomistic tunnel is more complex than for the
cylinder tunnel. There appears a significant number
of conformations of low $N_\mathrm{out}$, including
those of $N_\mathrm{out}=0$, at the atomistic but the cylinder tunnel. We have
checked that the trajectories that did not end with a successful escape after a
long time compared to $t_\mathrm{med}$ are associated with conformations of
$N_\mathrm{out}=0$. These conformations are identified as kinetic traps in the
escape process. 

Figure \ref{fig:trap} shows several trapped conformations obtained at
$T=0.75~\epsilon/k_B$ for GB1 at the atomistic tunnel. The number of native
contacts $N_c$ in these conformations is different but all of them have the
$\alpha$-helix and at least one
$\beta$-hairpin formed. The conformation shown
in Fig. \ref{fig:trap}(c) also has a partial tertiary structure established
by contacts between the $\alpha$-helix and the N-terminal $\beta$-hairpin. 
These conformations did not appear at the cylinder tunnel, indicating that
the irregular shape of the atomistic tunnel allows for and makes the
formation of trapped conformations more easy inside the tunnel.
Note that at the physiological temperature $T=0.85~\epsilon/k_B$, there
were no kinetic traps. This can be understood as due to two reasons: the faster
diffusion at this temperature helps the protein to avoid trapped conformations,
and the larger thermal fluctuations help the protein to get
out from the traps.
The trapping at an atomistic ribosome tunnel and the alleviation of
trapping by increased thermal fluctuations have been also observed for the
chymotrypsin inhibitor 2 (CI2) protein in an early simulation study by Elcock
\cite{Elcock2006}. Our results here for GB1 are consistent with that previous
work.

\begin{figure}
\includegraphics[width=8.5cm]{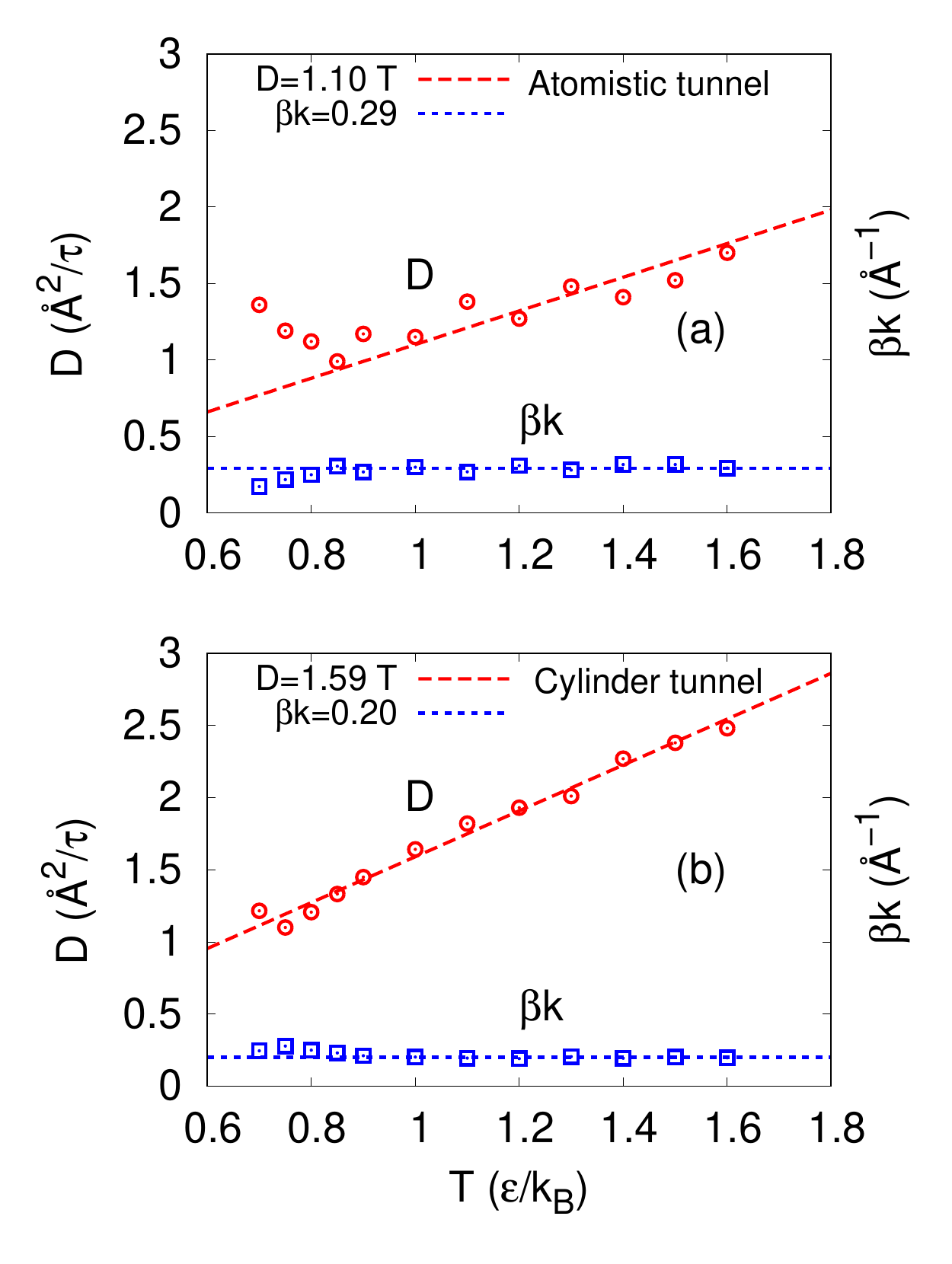}
\caption{Dependence of the diffusion constant $D$ (circles) and the parameter
$\beta k$ (squares) on temperature for protein GB1 at the atomistic tunnel (a)
and at the cylinder tunnel of $L=72$~{\AA} and $d=16$~{\AA} (b). The data are
obtained by fitting the simulated escape time distribution to Eq.
(\ref{eq:gt}).  Straight lines show fits of the simulation data for $T \geq
0.85~\epsilon/k_B$
to a linear dependence on temperature in the case of $D$ (dashed)
and to a constant value in the case of $\beta k$ (dotted).
The functions of the fits are $D=1.1\,T$ and $\beta k = 0.29$~{\AA}$^{-1}$ in
(a) and $D=1.59\,T$ and $\beta k = 0.2$~{\AA}$^{-1}$  in (b).
} 
\label{fig:Dbeta}
\end{figure}

It is interesting now to compare the protein diffusion properties at the two
tunnels using the diffusion model. Figure \ref{fig:Dbeta} shows the values of
the diffusion constant $D$ and the potential slope $\beta k$ obtained by
fitting the escape time
histograms at various temperatures to the diffusion model, wherein it can be
seen that for both the tunnels $D$ appears to increase linearly
with temperature whereas $\beta k$ tends to adopt a constant value. The linear
dependence of $D$ on temperature agrees with that of an ideal Brownian
particle. The atomistic tunnel, however, yields a lower $D$ and a higher
$\beta k$ than the cylinder tunnel at intermediate and high temperatures
($T \geq 0.85~\epsilon/k_B$),
while the average diffusion speed given by $D \beta k$ is almost the same
for the two tunnels. We have checked that at these temperatures, the escape
time distributions at the atomistic tunnel are slightly narrower than at the
cylinder tunnel.
At low temperatures ($T < 0.85~\epsilon/k_B$), both $D$ and $\beta k$ for the
atomistic tunnel deviate significantly from the average trends due to the
impact of kinetic trapping. Also at low temperatures, the escape time
distribution for the atomistic tunnel becomes broader than for the cylinder
tunnel.

\begin{figure}
\center
\includegraphics[width=8.5cm]{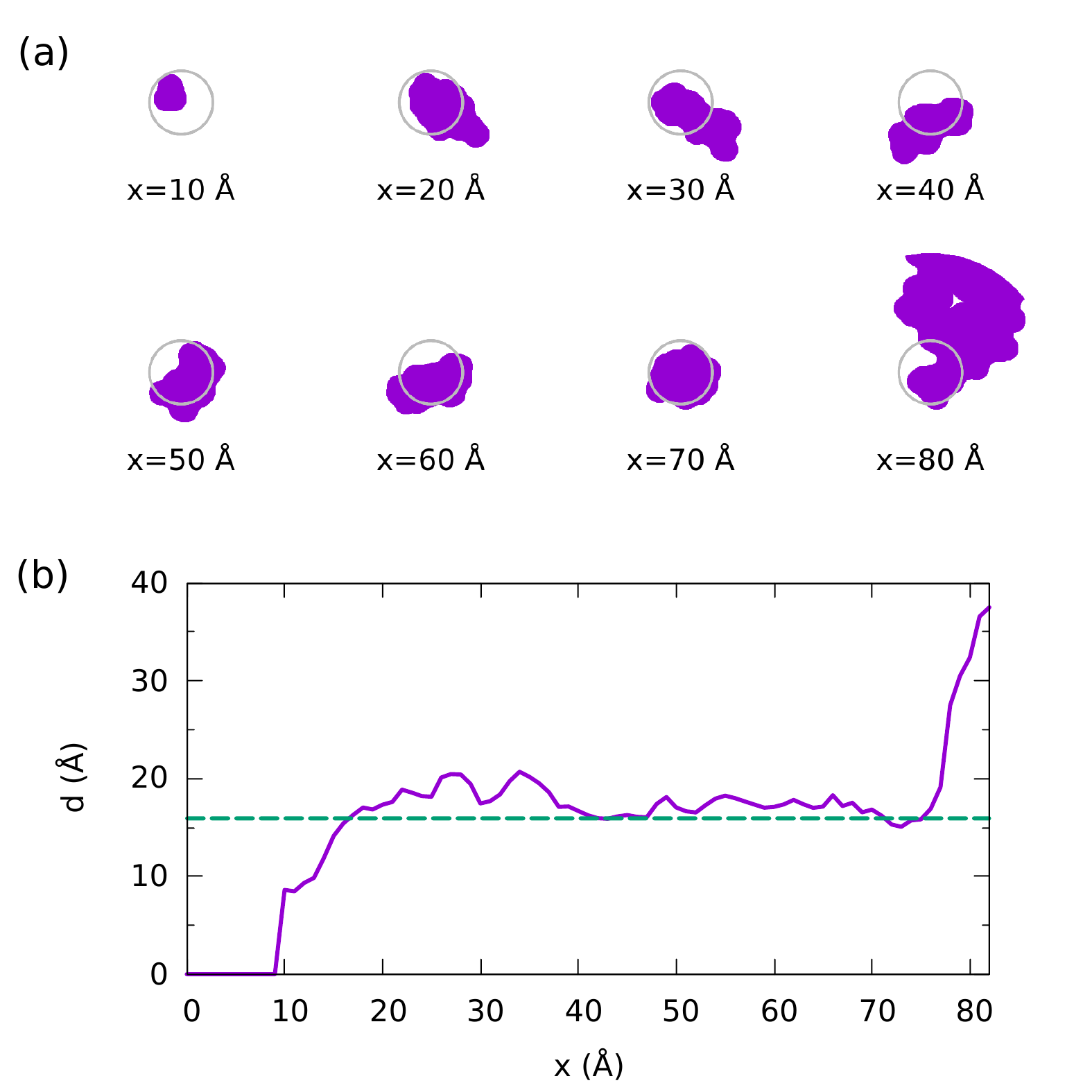}
\caption{(a) Inner cross-sections (dark color) of the atomistic tunnel
obtained by using a probe sphere of radius $R=3$~{\AA} at various
position $x$ along the tunnel axis as indicated.
For comparison, the cross-section of an equivalent cylinder tunnel of diameter
16~{\AA} is also shown (gray circle).
(b) Dependence of the effective diameter $d$ of the atomistic tunnel (solid) on
$x$. For a given position $x$, $d$ is calculated as $d=2\sqrt{S/\pi}$, where
$S$ is the area of the tunnel's cross-section. Horizontal dashed line indicates
the constant diameter $d=16$~{\AA} of the equivalent cylinder tunnel.
} \label{fig:diameter}
\end{figure}

To better understand the differences between the two tunnels,
we sought for a quantitative comparison between the shapes of the atomistic
tunnel and the cylinder tunnel. To that extent, we have calculated the area $S$
of the inner cross-section of the atomistic tunnel as a function of the position
$x$ along the tunnel axis using a probe sphere. The effective diameter of the
tunnel at a given position then was calculated as $d=2\sqrt{S/\pi}$. Figure
\ref{fig:diameter}(a) shows that the shape of the tunnel's cross-section varies
strongly with $x$. It is typically not circular and
significantly deviates from that of the equivalent cylinder tunnel. The tunnel
is also quite narrow near the PTC and becomes much wider at the exit port.
Figure \ref{fig:diameter}(b)
shows that the effective diameter $d$ of the atomistic tunnel varies, but not
too strongly, between 15 and 20~{\AA}, for the positions of
$x$ between 15 and 75~{\AA}. Notice that for these positions, the diameter
$d=16$~{\AA} of the equivalent cylinder tunnel lies within the variation range
of the atomistic tunnel's diameter but is near the lower bound of this range.
This can be understood as the irregular shape of the atomistic tunnel's
cross-sections makes it effectively smaller for nascent proteins.

The present model of the atomistic tunnel neglects the presence of amino
acid side-chains. We have checked that by considering all the heavy atoms of
ribosomal proteins in the tunnel model while keeping the C$_\alpha$-only
representation for GB1, the escape time distribution at $T=0.85~\epsilon/k_B$
changes only slightly with a small shift towards smaller values, but the 
escape probability reaches only about 95\% at the time of 1000~$\tau$
(Fig.~S2 of the supplementary material). We also found that due to
the increased roughness of the tunnel surface, the all-heavy-atom tunnel
requires 
a longer growth time of the protein, with $t_g = 400~\tau$ per amino acid,
to obtain converged properties of fully translated protein conformations. The
results indicate that amino acid side-chains may have a relatively small but
detrimental effect on the escape process. A
proper consideration of this effect, however, would need models that include
side-chain representations for both ribosomal and nascent proteins and allow
the degrees freedom of side-chain rotation. It can be expected that while
the side-chain exclude volumes cause some obstruction to the escape process,
the freedom of side-chain rotation can make the escape somewhat easier.

\subsection{Dependence of escape time on protein}

\begin{figure*}
\includegraphics[width=17cm]{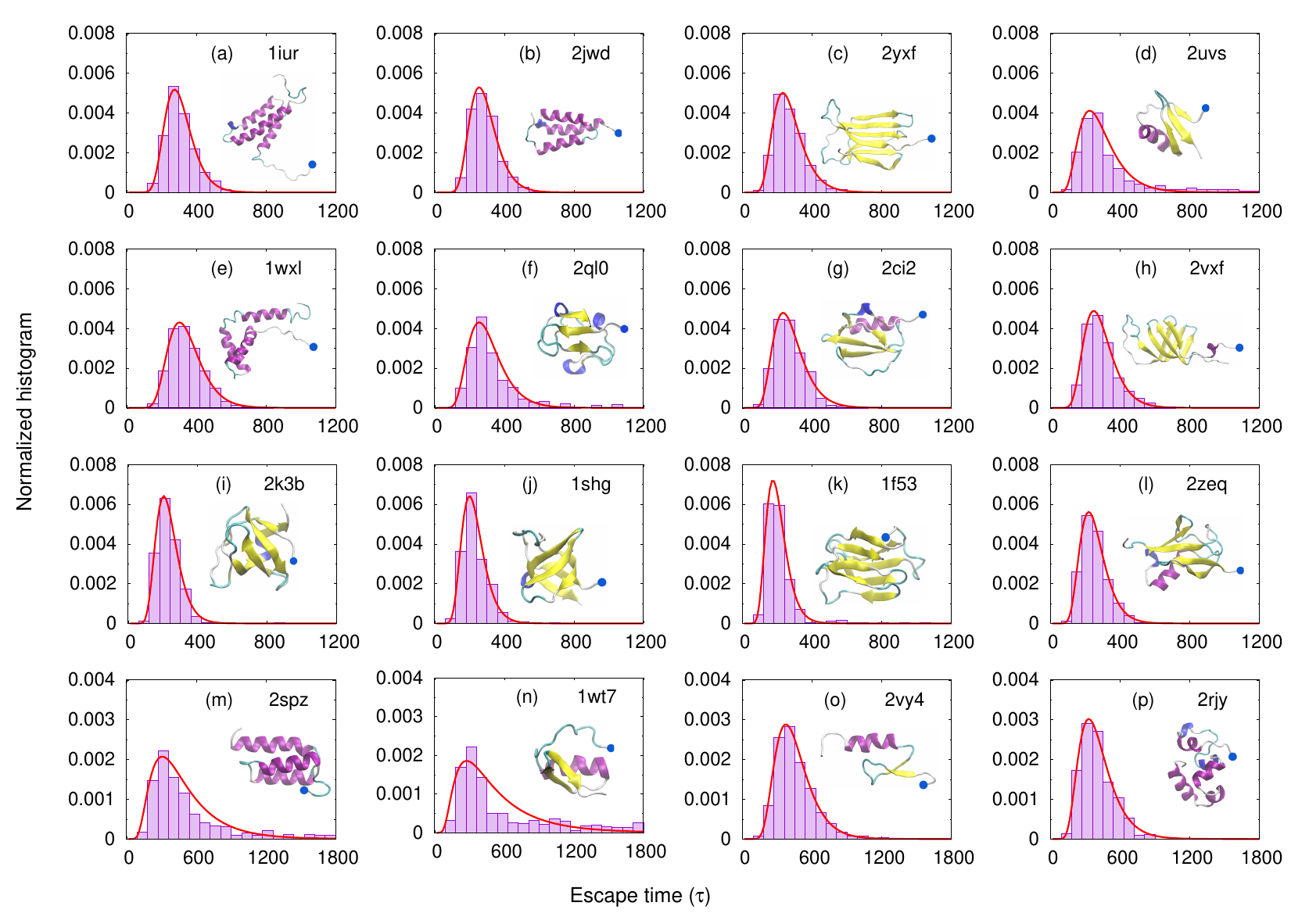}
\caption{Distribution of the escape time at the atomistic tunnel at
$T=0.85~\epsilon/k_B$ for 16 small single-domain proteins (not including GB1)
named by their PDB codes as 1iur (a), 2jwd (b), 2yxf (c), 2uvs (d), 
1wxl (e), 2ql0 (f), 2ci2 (g), 2vxf (h), 2k3b (i), 1shg (j), 1f53 (k),
2zeq (l), 2spz (m), 1wt7 (n), 2vy4 (o), 2rjy (p).
The PDB code and the native state of each protein are shown inside the panels
with the N-terminus indicated by a blue ball. 
In each panel, a normalized histogram of the escape times obtained from
simulations (boxes) is fitted to the diffusion model (solid line). The
fractions of non-escaped
trajectories at the largest time in
the histograms are 0\% (1iur), 1\% (2jwd), 0.3\% (2yxf), 4.6\% (2uvs), 0\% (1wxl),
5.6\% (2ql0), 0.6\% (2ci2), 0.2\% (2vxf), 1.2\% (2k3b), 1.2\% (1shg),
4\% (1f53), 0.6\% (2zeq), 10.9\% (2spz), 13.7\% (1wt7), 1\% (2vy4),
and 2.6\% (2rjy).
}
\label{fig:16prot}
\end{figure*}

In order to study the dependence of the escape time on protein, apart from
GB1 we selected additional 16 single domain proteins of lengths between 37 and
99 residues and belonging to different classes of all-$\alpha$, all-$\beta$ and
$\alpha/\beta$ proteins, and carried out simulations for these proteins. 
We consider only the atomistic tunnel and the simulation temperature
$T=0.85~\epsilon/k_B$. Figure \ref{fig:16prot} shows that the histograms of the
escape times for the selected proteins are quite similar in the overall shape
and the range of most probable values. The peak positions of the histograms
vary but within the same order of magnitude. The distribution width is the most
narrow for the all-$\beta$ proteins. For most of the proteins, the histogram
can be fitted quite well to the distribution function given by Eq.
(\ref{eq:gt}) of the diffusion model. For a few proteins, i.e. the ones with
PDB codes 2spz and 1wt7, the agreement with the diffusion model is worse
than for others with the appearance of a thick tail of large escape times
(Fig. \ref{fig:16prot}(m and n)). For most of the proteins, we observed 
trajectories with trapped conformations of $N_\mathrm{out}=0$. The fraction of
non-escaped trajectories at the largest time in the histograms shown in Fig.
\ref{fig:16prot} (1200~$\tau$ or 1800~$\tau$) is below 2\% for 10 out of 16
proteins (see the caption of Fig.~\ref{fig:16prot}). At a larger time of
8000~$\tau$, this fraction falls
below 2\% for all proteins except 2spz, for which this fraction remains at
9.7\%. Thus, most proteins can escape efficiently. We have checked that the
trapped conformations of 2spz typically have a two-helix bundle formed within
the tunnel, resulting in an increased difficulty for its escape.

\begin{figure*}
\includegraphics[width=16cm]{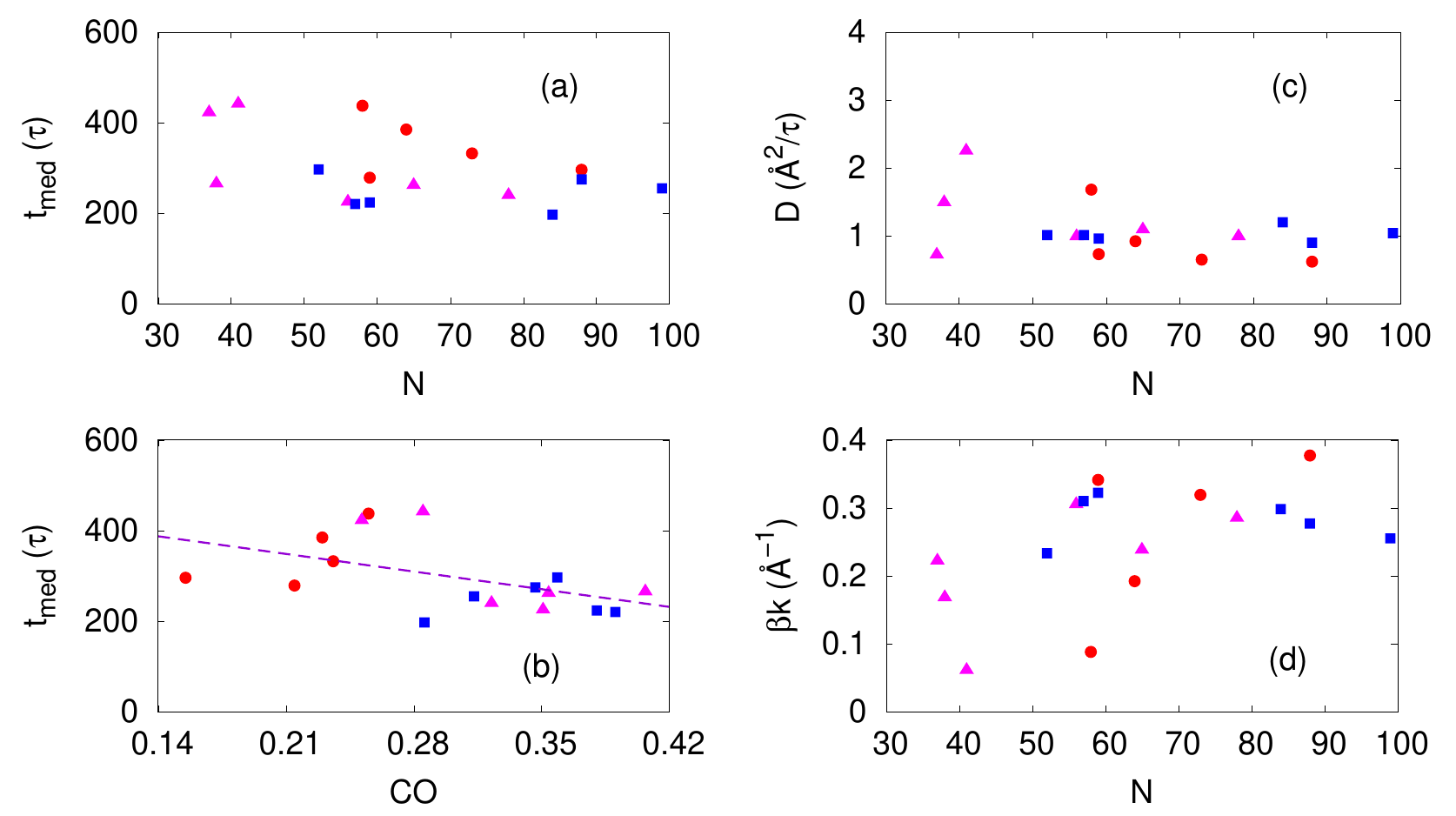}
\caption{(a and b) The median escape time $t_\mathrm{med}$ plotted against the
protein length $N$ (a) and the relative contact order CO (b) for GB1 and 16
proteins considered in Fig. \ref{fig:16prot}. The values of $t_\mathrm{med}$
are obtained from simulations with the atomistic tunnel at
$T=0.85~\epsilon/k_B$.  The point type indicates the protein class, i.e.
all-$\alpha$ (circles), all-$\beta$
(squares) and $\alpha/\beta$ (triangles).
(c and d) The diffusion constant $D$ (c) and the potential parameter $\beta k$
(d) plotted against the protein length $N$ for 17 proteins considered in (a).
$D$ and $\beta k$ are obtained by fitting the simulated escape time
distribution to the diffusion model.
}
\label{fig:tescL}
\end{figure*}

Figure \ref{fig:tescL}(a) shows the median escape time $t_\mathrm{med}$ as a
function of the chain length $N$ for all 17 proteins considered including GB1.
It can be seen that $t_\mathrm{med}$ is found in the range from 200~$\tau$ to
500~$\tau$ and does not seem to depend on $N$.  However, the variation of the
escape times among the proteins decreases with $N$.
Figure \ref{fig:tescL}(b) plots $t_\mathrm{med}$ against a topological
parameter of the protein native state, the relative contact order ($CO$)
\cite{Plaxco1998}. It shows an weak but visible trend that
$t_\mathrm{med}$ decreases with $CO$. From the 
types of data points shown in Fig. \ref{fig:tescL}(b) one can also see that the
all-$\beta$ proteins on averages escape faster than the all-$\alpha$ proteins,
whereas the $\alpha/\beta$ proteins can escape either as fast or as slow as the
two other groups. From the fits of the simulated escape time distribution to
the diffusion model, we obtained the values of $D$ and $\beta k$ for 17
proteins.
Figures \ref{fig:tescL}(c) and \ref{fig:tescL}(d) show that both $D$ and $\beta
k$ vary strongly among the proteins, but like for $t_\mathrm{med}$, the
variation decreases with
$N$ indicating a kind of convergence. It can be also noticed that the strongest
variation belongs to the $\alpha/\beta$ proteins, whereas the weakest belongs
to the $\beta$ proteins. It can be expected that proteins of length
$N>100$ have similar $t_\mathrm{med}$ and diffusion properties at the tunnel to
the ones of length $70<N<100$.

\begin{figure*}
\includegraphics[width=17cm]{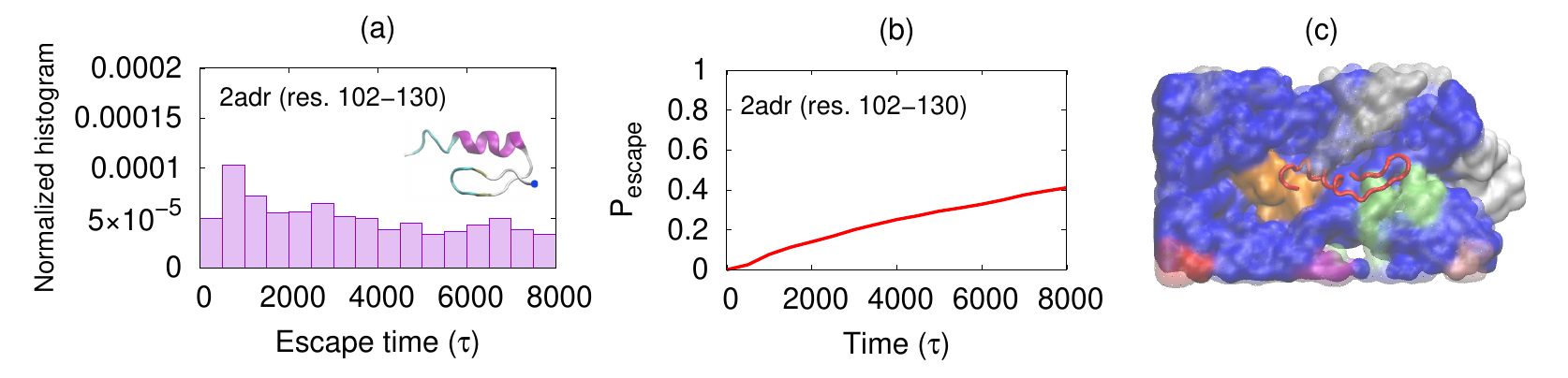}
\caption{(a) Histogram of the escape times at $T=0.85~\epsilon/k_B$ obtained
from simulations at the atomistic tunnel for a
29-residue zinc-finger domain of ADR1 protein (pdb code: 2adr, res. 102-130)
with the native structure of the domain shown as inset. (b) Dependence of
the escape probability, $P_\mathrm{escape}$, on time for the system considered
in (a). (c) A typical trapped conformation of the domain inside the tunnel.
}
\label{fig:zinc}
\end{figure*}

It is interesting that all the proteins considered are able to escape 
from the atomistic tunnel successfully including the CHHC zinc-finger domain
(2vy4) which has the length of only 37 residues. In order to check whether 
a smaller zinc-finger domain can escape, we carried out the simulations for
a 29-residue domain (res. 102-130) of ADR1 protein with the pdb code 2adr. 
Figure \ref{fig:zinc} shows that this zinc-finger domain poorly escapes the
tunnel with only 40\% of the trajectories being able to escape after a long
simulation time of 8000~$\tau$. A projected median escape time for this domain
would be at least about 50 times larger than for GB1.
Furthermore, the simulated escape time
distribution as shown Fig. \ref{fig:zinc}(a) cannot be fitted to the diffusion
model. Fig.  \ref{fig:zinc}(c) shows that a typical trapped conformation of
this zinc-finger domain has the $\alpha$-helix and the N-terminal
$\beta$-hairpin formed and it is found deeply within the tunnel. 
The partial folding observed here is consistent with a recent
experiment which indicates that the 2adr zinc-finger domain
can fold completely within the tunnel \cite{Nilsson2015}.
The difference
between the zinc-finger domain of 2adr and that of 2vy4 is that the latter is 8
residues longer and has a more developed $\beta$-hairpin with a coil-like
flagging tail near the N-terminus. This flagging tail makes the $\beta$-hairpin
formation inside the tunnel more difficult allowing the N-terminus to reach out
of the tunnel. It is suggested that the behavior of the 29-residue zinc-finger
domain is similar to the diffusion with $k=0$ in the diffusion model. A
protein trapped entirely inside the tunnel would feel no free energy
gradient and therefore has no indication on which direction to diffuse. It 
can be expected that the cross-over tunnel length \cite{Thuy2018} for the
29-residue zinc-finger domain is significantly shorter than the real length of
the exit tunnel and therefore the protein is found in the slow diffusion
regime.

\subsection{Dependence of escape time on friction coefficient}

\begin{figure}
\includegraphics[width=8.5cm]{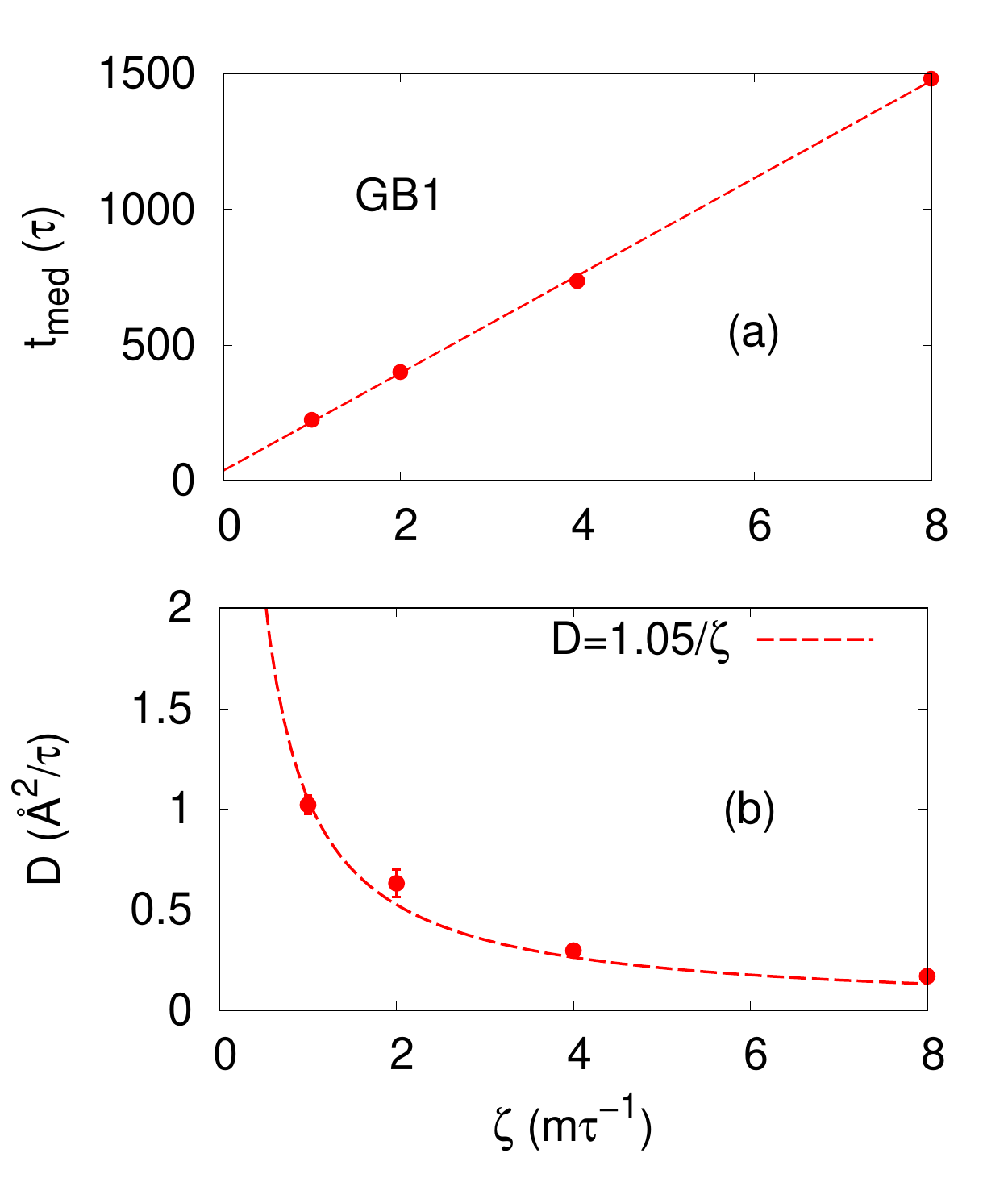}
\caption{Dependence of the median escape time $t_\mathrm{med}$ (a)
and the diffusion constant $D$ (b) on the friction coefficient $\zeta$
for protein GB1 at $T=0.85~\epsilon/k_B$. In (a) the dependence
is fitted by a linear function, $t_\mathrm{med} = 179.42\,\zeta + 37.62$
(dashed). In (b) the dependence is fitted by the function $D=1.05/\zeta$
(dashed).}
\label{fig:zeta}
\end{figure}

This subsection is relevant only to the methodology used in the study but it
helps to better interpret the previous results. All the simulations in the
previous subsections were done with the friction coefficient $\zeta =
1~m\tau^{-1}$ for amino acids.  This value of $\zeta$ may not be realistic for
real proteins inside cells. Thus, we ask how the escape time would depend on
$\zeta$ and whether one can extrapolate this dependence to obtain the real
escape time.  For this purpose we carried out additional simulations for GB1
with $\zeta=2$, 4, and 8~$m\tau^{-1}$ at the atomistic tunnel with
$T=0.85~\epsilon/k_B$. In these simulations, because the increased
friction slows down the dynamics, the growth time per
amino acid $t_g$ was also increased to 200, 400 and 800~$\tau$, respectively, for
the given values of $\zeta$.
Fig. \ref{fig:zeta}(a) shows that the median escape time
$t_\mathrm{med}$ is an almost perfect linear function of $\zeta$.  This linear
dependence  indicates that the simulation results are in the overdamped (large
friction) regime.  Fig. \ref{fig:zeta}(b) shows that the diffusion constant $D$
of the escaping protein, obtained by fitting the simulated escape time
distribution to the diffusion model, decreases with $\zeta$ like $D \sim
\zeta^{-1}$. Together with the approximate linear dependence of $D$ on
temperature shown Fig. \ref{fig:Dbeta}, one finds a complete consistency with
the Einstein's relation $D = k_B T/\zeta^*$ where $\zeta^*$ is the friction
coefficient of a Brownian particle.  Thus, protein at the tunnel behaves very
much like a Brownian particle if one assumes that the quantity $\zeta^*$ is
proportional to $\zeta$ and plays the role of an effective friction coefficient
of the whole protein.

Given that $\sigma=5$~{\AA}, $m = 110$~g/mol, and $\epsilon \approx
0.7$~kcal/mol, the time unit in our simulation can be calculated as $\tau =
\sqrt{m \sigma^2/\epsilon} \approx 3$~ps.  The simulation's friction unit
$m\tau^{-1} \approx 6 \times 10^{-11}$~g\,s$^{-1}$.  The realistic friction
coefficient of amino acid in water can be obtained from the Stokes law,
$\zeta_\mathrm{water} = 6\pi \eta \sigma$, where $\eta=0.01$~Poise is the
viscosity of water at 25$^\circ$C. One obtains $\zeta_\mathrm{water} \approx
9.4 \times 10^{-9}$~g\,s$^{-1} \approx 157~m\tau^{-1}$.  By extrapolating the
linear dependence in Fig. \ref{fig:zeta}(a) one finds that at $\zeta =
\zeta_\mathrm{water}$ the median escape time for GB1 is $t_\mathrm{med} \approx
3 \times 10^4~\tau \approx 90$~ns.  This time appears to be too short for large
scale motion like the protein escape.

Veitshans {\it et al.} \cite{Veitshans1997} suggested that at high friction,
inertial terms in the Langevin equation are irrelevant, and the natural time
unit is
$\tau_H = \zeta \sigma^2/k_B T$.  For water at room temperature, a direct
calculation from the last formula gives $\tau_H \approx 0.6$~ns. With
some scaling factor, Veitshans
{\it et al.} \cite{Veitshans1997} estimated that $\tau_H \approx 3$~ns.
With these revised time units, the theoretical estimate of $t_\mathrm{med}$ is
either 18~$\mu$s or 90~$\mu$s.  The experimental
refolding time of GB1 at neutral pH is about 1 ms \cite{Alexander1992}.  We have
checked that within the same Go-like model 
at $T=0.85~\epsilon/k_B$, the median refolding time is about 50\% larger
than $t_\mathrm{med}$.
Thus, the model prediction of the refolding time is smaller but within the
same order of magnitude as the experimental value given that some uncertainties
are associated with the estimates.
With the above estimates, and given the results of the previous section,
it can be expected that the escape times of single-domain proteins are
of the order of 0.1 ms, i.e. in the sub-millisecond scale.

\section{Conclusion}

There are several remarks we would like to mention for the conclusion.
First, the shape of the ribosomal exit tunnel appears to cause increased
difficulty for nascent proteins to escape compared to a smooth cylinder tunnel. 
This difficulty is reflected by the appearance of kinetic traps in the 
escape pathways leading to lengthened escape times. We have shown that the
trapped conformations are completely located inside the tunnel and usually have a
significant development of tertiary structure. The formation of tertiary
structure elements inside the tunnel correlates with the modulated shape of the
ribosome tunnel, which has some narrow parts but also some wider parts which
can hold a tertiary unit. In contrast, the equivalent cylinder tunnel
of 13.5~{\AA} diameter does not allow tertiary structure formation and yields
no kinetic trapping.
Second, thermal fluctuations are important for the escape of nascent proteins.
We have shown that for GB1, a significant fraction of escape
trajectories get trapped at $T=0.75~\epsilon/k_B$, but not at the physiological
temperature $T=0.85~\epsilon/k_B$. Interestingly, at the latter temperature,
almost all of the 17 single-domain proteins considered are able to escape
efficiently, even the smallest one, the 37-residue 2vy4. 
Note that the trapping arises solely due to the folding of a protein
within the tunnel, thus it depends on temperature. At high temperatures,
folding is slow and diffusion is fast, therefore a protein would have a 
low probability of getting trapped
before escaping from the tunnel. At low temperatures, folding is fast while
diffusion is slow, the trapping probability is increased. On the other hand,
a protein would get out from a trap easier at a higher temperature.

Third, if a protein or peptide is too small it cannot escape efficiently from
the tunnel. The example of the 29-residue zinc-finger domain of 2adr
shows that the protein is severely trapped inside the tunnel with 
the median escape time about of two orders of magnitude larger than that
of GB1. The trapped protein is not guided by a potential gradient towards the
escape direction. This example reflects a relation between the protein size
and a cross-over tunnel length for efficient diffusion, as
predicted by our previous study with the cylinder tunnel \cite{Thuy2018}.
Forth, the escape time of single-domain proteins weakly depends on the native
state topology and is almost independent of the protein size. Our
model predicts that the protein escape time at the ribosome
tunnel is of the order of 0.1 ms. The latter is much shorter than
the time needed by the ribosome to translate one codon (tens of milliseconds),
therefore not allowing nascent proteins to jam the ribosome tunnel.

One may ask to what extent hydrophobic and electrostatic interactions of a
nascent protein with the ribosome exit tunnel can alter the above obtained
results. It is well-known that the tunnel's wall formed by the ribosomal RNA
is negatively charged.  We found that for {\it H. marismortui}'s ribosome, the
tunnel's inner surface with $x<82~${\AA} has only four
hydrophobic side-chains that are clearly exposed within the tunnel, belonging
to Phe61 of protein L4, Met130 of protein L22, and Met26 and Leu27 
of protein L38, and about 10 exposed charged amino acid side-chains.  
These statistics indicate that the effect of hydrophobic interaction on the
escape process can be considerably small whereas the Coulomb interaction may
have a strong effect on nascent chains.  However, if the total charge of a
nascent protein is
neutral, the electrostatic forces on the protein's positive and negative
charges may cancel out each other. Thus, it is reasonable to expect that the
energetic interactions of nascent proteins with the tunnel can lead to specific
changes in the escape behavior for individual proteins, but on an average they
give only higher-order corrections to what obtained with excluded volume
interaction.
 
Finally, like for the cylinder tunnel, it is found that the escape time
distribution at the atomistic tunnel for various proteins follows very well the
one-dimensional diffusion model of a drifting Brownian particle.
This consistent finding suggests that the protein escape at the ribosome tunnel
may have been designed by Nature to be simple, efficient and predictable for
the smooth functioning of the ribosome. This result also proves the usefulness
of using simple stochastic models to understand complex dynamics of
biomolecules.

\section*{Supplementary Material}

See supplementary material 
for the dependences of the median escape time and the mean escape time
on temperature for protein GB1 at the cylinder tunnel of length $L=82$~{\AA} and
diameter $d=13.5$~{\AA}, and
for the histogram of the escape time and the time dependences of the escape
probability and the probability of the C-terminal $\beta$-hairpin formation
inside the tunnel for protein GB1 at a tunnel model that
considers all the heavy atoms of the ribosomal RNA and the ribosomal proteins.

\begin{acknowledgments}
This research is funded by Vietnam National Foundation for Science and
Technology Development (NAFOSTED) under grant number 103.01-2019.363.  
T.X.H. also acknowledges the support of the International
Centre for Physics at the Institute of Physics, VAST under grant number
ICP.2020.05.
We thank the VNU Key Laboratory of Multiscale Simulation of Complex
Systems for the occasional use of their high performance computer.
\end{acknowledgments}

\section*{Data availability statement}

The data that support the findings of this study are available from the
corresponding author upon reasonable request.


%

\newpage

\setcounter{equation}{0}
\renewcommand\theequation{S\arabic{equation}}

\setcounter{figure}{0}
\renewcommand\thefigure{S\arabic{figure}}

\setcounter{table}{0}
\setcounter{page}{1}
\renewcommand{\bibnumfmt}[1]{[S#1]}
\renewcommand{\citenumfont}[1]{[S#1]}

\onecolumngrid

\makeatletter

\section*{Supplementary Material for:
``Protein escape at the ribosomal exit tunnel: effect of the tunnel shape'',
P.T. Bui and T.X. Hoang}

\begin{figure}[!ht]
\includegraphics[width=8cm]{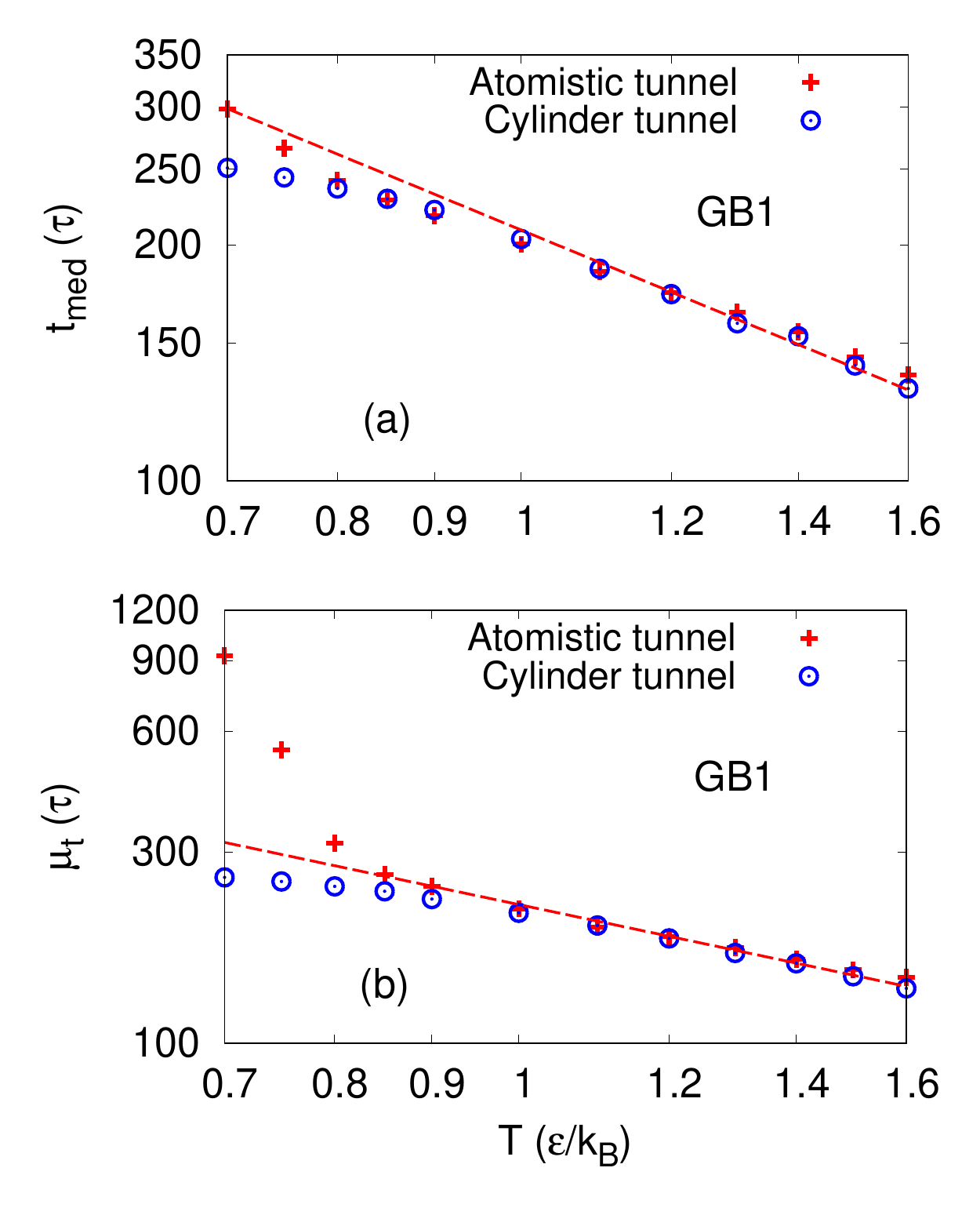}
\caption{Same as for Figure 2 but for a comparison 
between the atomistic
tunnel (crosses) with a cylinder tunnel of length
$L=82$~{\AA} and diameter $d=13.5$~{\AA} (circles).}
\label{fig:s1}
\end{figure}

\begin{figure}[!ht]
\includegraphics[width=15cm]{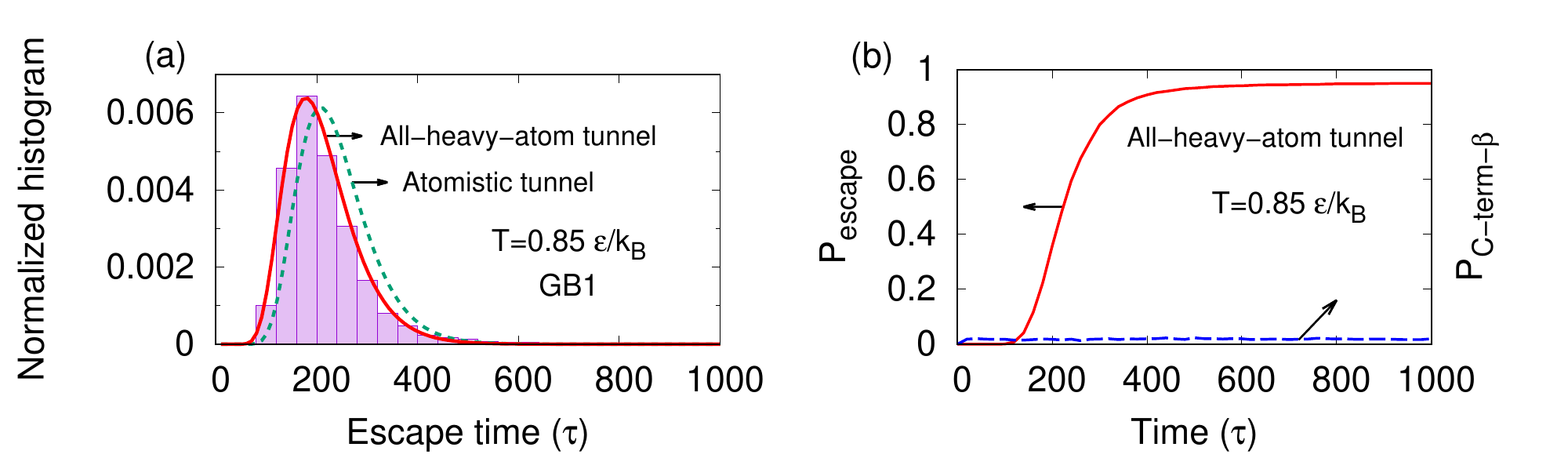}
\caption{Same as for panels (a) and (b) of Figure 7 but for a tunnel
model that consists of all the heavy atoms of the ribosomal RNA and the
ribosomal proteins. The GB1 protein is still considered in the C$_\alpha$-based
Go-like model. The fit of the simulation data to the diffusion model 
for the atomistic tunnel (dotted) is shown for comparison with the fit
for the all-heavy-atom tunnel (solid).
Compared to the atomistic tunnel, the all-heavy-atom tunnel requires a longer
growth time of the protein to obtain converged properties of fully translated
protein conformations in terms of number of native contacts and radius of
gyration. For the results shown in the figure, the protein is
grown with the growth time $t_g=400~\tau$ per amino acid. A longer
growth time leads to similar results.
}
\label{fig:s2}
\end{figure}

\end{document}